\documentclass[a4paper,11pt,noarxiv]{quantumarticle}
\pdfoutput=1
\usepackage[utf8]{inputenc}
\usepackage[T1]{fontenc}
\usepackage{amsmath,amssymb,amsfonts,amsthm}
\usepackage{braket}
\usepackage{graphicx}
\usepackage{enumitem}
\usepackage{booktabs}
\usepackage{multirow}
\usepackage[qm]{qcircuit}
\usepackage{tikz}
\usepackage{adjustbox}
\usetikzlibrary{decorations.pathmorphing,decorations.pathreplacing,calc,arrows.meta,positioning,shapes.geometric,fit,backgrounds}
\usepackage{hyperref}
\usepackage{xcolor}

\DeclareMathOperator{\tr}{tr}
\DeclareMathOperator{\Stab}{Stab}

\newtheorem{corollary}{Corollary}

\newtheorem{definition}{Definition}
\newtheorem{proposition}{Proposition}

\newcommand{\Crank}{C_{\mathrm{rank}}}
\newcommand{\Lrank}{L^{*}_{\mathrm{rank}}}

\newtheorem{theorem}{Theorem}
\theoremstyle{remark}

\newcommand{\R}{\mathbb{R}}
\newcommand{\Z}{\mathbb{Z}}
\newcommand{\C}{\mathbb{C}}

\newcommand{\Fprod}{\mathcal{F}_{\mathrm{prod}}}

\newcommand{\manifold}{\mathcal{M}}
\newcommand{\supp}{\mathrm{supp}}
\newcommand{\rank}{\operatorname{rank}}
\newcommand{\nobs}{n_{\mathrm{obs}}}

\begin{document}
\title{Data informed Ansatz Selection for Variational Quantum Circuits}

\author[1]{Manav Seksaria}
\email{manavs@ee.iitm.ac.in}

\author[3]{Advith Desu}
\email{jaisaiadvith\_desu@mymail.sutd.edu.sg}

\author[1,2]{Anil Prabhakar}
\email{anilpr@ee.iitm.ac.in}

\affiliation[1]{Center for Quantum Information, Communication and Computing, Indian Institute of Technology Madras, Chennai, India}
\affiliation[2]{Department of Electrical Engineering, Indian Institute of Technology Madras, Chennai, India}
\affiliation[3]{Engineering Systems and Design, Singapore University of Technology and Design, Singapore}

\date{\today}

\maketitle

\begin{abstract}
Choosing a variational quantum circuit (VQC) usually means searching over ans\"atze and depths, and training each candidate. We give three rules that fix these choices from the data and the gate set alone, before any training. First, a circuit without entangling gates has a rank-one Fourier coefficient tensor on every subset of qubits. A singular value decomposition of a dataset's Fourier coefficients therefore indicates whether entanglement is needed, which qubits it connects, and a lower bound on the error of omitting it. Second, the rank of the quantum Fisher information matrix grows by $p_0$ per layer, after a first-layer deficit $d_0$, until it reaches a ceiling set by the dynamical Lie algebra. This law fixes the depth. Third, reading out all $2^N - 1$ Pauli-$Z$ products from one batch of shots resolves half of the state's directions, the most any single-basis readout can. A single-qubit readout resolves an exponentially small fraction. For the Madelon dataset, the rules prescribe $7$ blocks of $3$ qubits at depth $3$: a $239$-parameter classifier with $78.5\%$ test accuracy, found without a grid search. Ablations confirm three predicted effects: shallower circuits, smaller blocks and dropping reuploading all lower accuracy.
\end{abstract}

\section{\label{sec:intro}Introduction}
Designing a variational quantum circuit (VQC) for a learning task starts today with a search over candidate ans\"atze and depths. The search is run by brute force or by a learning-based method such as quantum architecture search, reinforcement learning, or Bayesian optimisation~\cite{martyniuk2024qas}. Either way, the search costs a lot of computation before any model is trained.

Circuits can be judged without training by analytic measures such as expressibility, entangling capability~\cite{sim2019}, and effective-dimension~\cite{abbas2021}. These describe what a given circuit can do. They do not say which circuit to choose, or at what depth. This work proposes analytical rules that fix several of these choices without running or training any circuit.

\begin{figure}[!htpb]
\centering
\begin{tikzpicture}[
  node distance=3mm and 3mm,
  box/.style={draw, rounded corners=1pt, minimum height=8mm, minimum width=12mm, align=center, font=\footnotesize},
  arr/.style={-{Stealth[length=1.7mm]}, thick},
  ann/.style={font=\scriptsize, align=center}
]
\node[box] (in)  {input\\$x$};
\node[box, right=of in]  (enc) {encode\\$U(x,\boldsymbol\theta)$};
\node[box, right=of enc] (st)  {state\\$\ket{\psi(x,\boldsymbol\theta)}$};
\node[box, right=of st]  (m)   {readout\\$\langle O\rangle$};
\node[box, right=of m]   (cls) {logit\\$f(x)$};
\draw[arr] (in)  to (enc);
\draw[arr] (enc) to (st);
\draw[arr] (st)  to (m);
\draw[arr] (m)   to (cls);
\node[ann, below=2mm of enc, xshift=-3.5mm] (a1) {Representability\\Sec.~\ref{sec:function-class}};
\node[ann, below=2mm of st]  (a2) {QFIM rank\\Sec.~\ref{sec:rank}};
\node[ann, below=2mm of m]   (a3) {Extraction\\Sec.~\ref{sec:extraction}};
\draw[densely dotted] (enc.south) to (enc.south |- a1.north);
\draw[densely dotted] (st.south)  to (a2.north);
\draw[densely dotted] (m.south)   to (a3.north);
\end{tikzpicture}
\caption{One consideration applies at each stage of the circuit's information path: representability at the encoding, QFIM rank at the state, extraction at the readout.}
\label{fig:pipeline}
\end{figure}
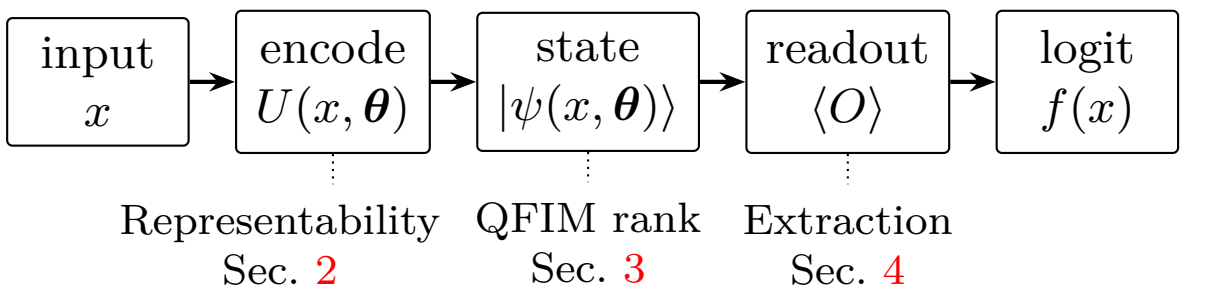

We focus on three considerations, each tied to a distinct stage of how a parameterised circuit converts inputs into classifications (Fig.~\ref{fig:pipeline}).

\textbf{Representability.} Bowles, Ahmed, and Schuld~\cite{bowles2024} benchmark a dozen quantum classifiers on six binary classification tasks. The variant with entangling gates removed often matches or beats its parent, and one cannot tell which is better without training both. The structure of the target labels decides whether a circuit needs entangling gates. We propose a diagnostic that runs before training. From a labelled dataset, it decides whether a product ansatz can represent the target and, if not, which qubits must be entangled (Sec.~\ref{sec:function-class}).

\textbf{QFIM rank.} The second consideration is how many of a circuit's parameters move the state in independent directions. The rank of the quantum Fisher information matrix (QFIM) counts them. We show that the rank grows linearly with depth until it reaches a ceiling set by the dynamical Lie algebra (DLA) of the gate set (Sec.~\ref{sec:rank}).

\textbf{Extraction.} The readout fixes how much of the state a VQC can use. The extraction ratio $\rho$ measures this (Sec.~\ref{sec:extraction}). Reading out all $2^N - 1$ Pauli-$Z$ products gives the largest $\rho$ of any single-basis readout, and one batch of computational-basis shots provides them all.

\section{Representability}
\label{sec:function-class}
A circuit without entangling gates can only compute functions whose Fourier coefficients factorise. This section proves that constraint and turns it into a test that decides, from the data alone, whether entanglement is needed.

\subsection{Motivation}

Entangling gates couple pairs of qubits and are expensive, but they reach functions that no arrangement of single-qubit gates can produce. For a given learning task, the design question is whether they are needed. A VQC without entangling gates is a product ansatz.

Our result builds on Schuld, Sweke, and Meyer~\cite{schuld2021}, who showed that a VQC computes a trigonometric polynomial whose frequencies are fixed by the data encoding. Their analysis says which Fourier modes a circuit can reach. Casas and Cervera-Lierta~\cite{casas2023} note that without entanglement the circuit computes a product of one-dimensional Fourier series. We make this exact as a rank-one condition on the coefficients of every subset of qubits. Together, the two results give a test, run on a labelled dataset, of whether entanglement is needed. Wiedmann et al.~\cite{wiedmann2025} also compare the Fourier transform of a dataset with the spectrum of candidate circuits to rank architectures. They do not test for entanglement.

\subsection{Setup and notation}
\label{subsec:setup}

We consider an $N$-qubit VQC acting on the initial state $\ket{0}^{\otimes N}$, which we write $\ket{0}$. The input $x = (x_0, x_1, \ldots, x_{N-1}) \in \R^N$ is encoded one coordinate per qubit: $x_q$ enters qubit $q \in [N] = \{0, 1, \ldots, N-1\}$ through the rotation $R_y(x_q) = \exp(-i x_q Y/2)$. The circuit has $L$ layers. Each layer applies this encoding to every qubit, then variational single-qubit rotations and, optionally, entangling gates. We write $\ket{\psi(x,\boldsymbol\theta)}$ for the resulting state, with $\boldsymbol\theta$ the vector of all variational angles.

The circuit's output is the expectation value of an observable $M$. We take $M$ to be diagonal in a product basis. Local rotations can be absorbed into the circuit, so this basis can be the computational one, and $M$ is a weighted sum of Pauli-$Z$ products. We label each product by a binary string $\alpha = (\alpha_0, \ldots, \alpha_{N-1}) \in \{0,1\}^N$, where $\alpha_q = 1$ marks the qubits on which $Z$ acts and $\alpha_q = 0$ the qubits on which the identity acts:
\begin{equation}
M = \sum_{\alpha \in \{0,1\}^N} w_\alpha \, Z^\alpha,
\qquad
Z^\alpha = \bigotimes_{q=0}^{N-1} Z_q^{\alpha_q},
\label{eq:observable}
\end{equation}
with real weights $w_\alpha \in \R$. Here $Z_q$ and $I_q$ are the Pauli-$Z$ and identity operators on qubit $q$, with $Z_q^0 = I_q$ and $Z_q^1 = Z_q$.

Writing $\langle Z^\alpha(x) \rangle = \bra{\psi(x,\boldsymbol\theta)} Z^\alpha \ket{\psi(x,\boldsymbol\theta)}$ for the expectation of $Z^\alpha$, the function computed by the circuit is
\begin{equation}
f(x) = \sum_{\alpha \in \{0,1\}^N} w_\alpha \, \langle Z^\alpha(x) \rangle.
\label{eq:fdef}
\end{equation}

Schuld, Sweke, and Meyer~\cite{schuld2021} showed that any function of the form of Eq.~\eqref{eq:fdef} is a trigonometric polynomial,
\begin{equation}
f(x) = \sum_{\ell \in \Omega} c(\ell) \, e^{i \ell \cdot x},
\qquad
\ell \cdot x = \sum_{q=0}^{N-1} \ell_q x_q,
\label{eq:fourier}
\end{equation}
where $c(\ell) \in \C$ is the Fourier coefficient of the mode at frequency $\ell = (\ell_0, \ldots, \ell_{N-1})$. The sum runs over the frequency support $\Omega \subseteq \Z^N$, the frequencies at which the circuit can have non-zero coefficients. The spectrum of the encoding generator fixes $\Omega$~\cite{schuld2021}. For the $R_y$ encoding used here, $\Omega \subseteq \{-L,\ldots,L\}^N$. The \emph{support} of a multi-index $\ell \in \Z^N$ is the set of qubits on which it is non-zero,
\begin{equation}
\supp(\ell) = \{q : \ell_q \neq 0\}.
\end{equation}

\subsection{Product factorisation}
\label{subsec:product-factorisation}

We now restrict to circuits with no entangling gates (product ans\"atze) and write $\Fprod$ for the family of functions they compute. With only single-qubit gates acting on $\ket{0}$, the state is a product at every stage of the circuit,
\begin{equation}
\ket{\psi(x,\boldsymbol\theta)} = \bigotimes_{q=0}^{N-1} \ket{\psi_q(x_q,\boldsymbol\theta_q)},
\end{equation}
where $\boldsymbol\theta_q$ holds the variational angles on qubit $q$, and each factor depends only on the local coordinate $x_q$.

The factorisation passes to the readout. Since $Z^\alpha$ is a tensor product of single-qubit operators, and the state is a tensor product of single-qubit states,
\begin{equation}
\langle Z^\alpha(x) \rangle = \prod_{q=0}^{N-1} f_q(x_q)^{\alpha_q}.
\label{eq:corr-factorise}
\end{equation}
Here, $f_q(x_q) = \bra{\psi_q(x_q,\theta_q)} Z \ket{\psi_q(x_q,\theta_q)} \in [-1, 1]$ is the single-qubit Pauli-$Z$ expectation. The convention $f_q^0 = 1$ mirrors $Z_q^0 = I$: qubits with $\alpha_q = 0$ contribute a factor of one. Substituting Eq.~\eqref{eq:corr-factorise} into Eq.~\eqref{eq:fdef} gives the explicit form of $\Fprod$,
\begin{equation}
\Fprod = \sum_{\alpha \in \{0,1\}^N} w_\alpha \prod_{q=0}^{N-1} f_q(x_q)^{\alpha_q}.
\label{eq:product-class}
\end{equation}

\subsection{Rank-one constraint}
\label{subsec:rank-one}

Applied to a single qubit, the expansion of Eq.~\eqref{eq:fourier} gives
\begin{equation}
f_q(x_q) = \sum_{\ell_q} c_q(\ell_q) e^{i \ell_q x_q},
\end{equation}
with coefficients $c_q(\ell_q) \in \C$ supported on $\{-L, \ldots, L\}$. These coefficients form the per-qubit Fourier vector $c_q$. Substituting into Eq.~\eqref{eq:product-class} and matching Fourier modes shows that the Fourier coefficients of $f$ supported on any fixed subset $S \subseteq [N]$ are products of per-qubit coefficients $c_q(\ell_q)$, up to a scalar that depends only on $S$.
For any $S \subseteq [N]$ with $|S| \geq 2$, collect the coefficients $\{c(\ell) : \supp(\ell) = S\}$ into the \emph{Fourier coefficient tensor} on $S$,
\begin{equation}
C^{(S)}_{(\ell_q)_{q \in S}} = c(\ell), \qquad \ell_q \in \Z \setminus \{0\} \text{ for } q \in S,
\end{equation}
an order-$|S|$ array with each axis of length $2L$ (indexed by $\{\pm 1, \ldots, \pm L\}$) for an $R_y$ encoding at depth $L$.

\begin{theorem}[Rank-one constraint]
\label{thm:rank-one-constraint}
For any $f \in \Fprod$ and any $S \subseteq [N]$ with $|S| \geq 2$, the Fourier coefficient tensor satisfies $\rank(C^{(S)}) \leq 1$.
\end{theorem}
\begin{proof}
Each term $w_\alpha\prod_{q\in\supp(\alpha)} f_q(x_q)$ of Eq.~\eqref{eq:product-class} is a product of univariate Fourier series in disjoint variables, so its coefficient at $\ell$ is $w_\alpha\prod_{q\in\supp(\alpha)} c_q(\ell_q)$~\cite{casas2023}. It contributes to $\supp(\ell) = S$ only if $\supp(\alpha)\supseteq S$, with the other qubits entering through their constant terms $c_{q'}(0)$. Summing over $\alpha$ gives $C^{(S)} = W_S\bigotimes_{q\in S} c_q|_{\neq 0}$ for a scalar $W_S$, a single outer product of complex vectors, so its rank over $\C$ is at most one.
\end{proof}

Here $\rank$ is taken over $\C$, since the coefficients are complex. For $|S| = 2$ it is the matrix rank of $C^{(S)}$. For $|S| \geq 3$, it is the tensor rank, the smallest number of outer products of complex vectors that sum to $C^{(S)}$.

Let $f^*$ be the target, with Fourier coefficients $c^*(\ell)$ and tensors $C^{*(S)}$.

\begin{corollary}[Unreachability witness]
\label{cor:unreachable}
If $\rank(C^{*(S)}) \geq 2$ for some $S$ with $|S| \geq 2$, then no product ansatz at any depth represents $f^*$ exactly: the qubits in $S$ must be entangled.
\end{corollary}

A generic trigonometric polynomial fails the rank-one constraint on at least one subset $S$. For example, $\cos(x_0 + x_1) = \cos x_0 \cos x_1 - \sin x_0 \sin x_1$ has Fourier coefficient matrix (in the real basis) $\mathrm{diag}(1,-1)$, of rank two. No product ansatz, at any depth, can produce $\cos(x_0+x_1)$.

When $f^* \notin \Fprod$, the same constraint gives a lower bound on the mean squared error (MSE) of any product ansatz. The circuit output is $2\pi$-periodic in every coordinate. We measure error in the squared $L^2$ norm on the torus $\mathbb{T}^N = (\R/2\pi\Z)^N$ with normalised measure, in which $\{e^{i\ell\cdot x}\}_{\ell\in\Z^N}$ is orthonormal. For a target $f^*$ with Fourier coefficient tensors $C^{*(S)}$, let
\begin{equation}
\rho_1(C^{*(S)}) := \min_{R : \rank(R) \leq 1} \|C^{*(S)} - R\|_F^2
\end{equation}
be its best rank-one approximation error. For $|S| = 2$, the best rank-one approximation is the leading term of the singular value decomposition~\cite{kolda2009}. Hence,
\begin{equation}
\rho_1(C^{*(S)}) \;=\; \sum_{k \geq 2} \sigma_k(C^{*(S)})^2,
\label{eq:rho1-svd}
\end{equation}
with $\sigma_1 \geq \sigma_2 \geq \cdots$ the singular values of $C^{*(S)}$. For $|S| \geq 3$, the same tail over any mode unfolding is a lower bound, since the mode-$n$ unfolding of a rank-one tensor is a rank-one matrix~\cite{kolda2009}.

\begin{proposition}[MSE lower bound]
\label{prop:mse-bound}
For any target $f^* \in L^2(\mathbb{T}^N)$ and any $f \in \Fprod$,
\begin{equation}
\|f^* - f\|_{L^2}^2 \;\geq\; \sum_{S \subseteq [N] : |S| \geq 2} \rho_1(C^{*(S)}).
\end{equation}
\end{proposition}
\begin{proof}
By Parseval's identity, $\|f^* - f\|_{L^2}^2 = \sum_{\ell} |c^*(\ell) - c(\ell)|^2$. We group the terms by the support $S$ of $\ell$. Terms with $|S| \leq 1$ are non-negative. For $|S| \geq 2$ the group equals $\|C^{*(S)} - C^{(S)}\|_F^2$, and Theorem~\ref{thm:rank-one-constraint} makes $C^{(S)}$ rank-one, so it is at least $\rho_1(C^{*(S)})$. Summing over $S$ gives the bound.
\end{proof}
The bound is computed from the target alone. It is not tight in general: one per-qubit Fourier vector is shared by every subset that contains the qubit, so the exact problem is a coupled low-rank approximation of the kind studied in collective matrix factorisation~\cite{singh2008}.

\subsection{Pre-training diagnostic}
\label{subsec:diagnostic}

Corollary~\ref{cor:unreachable} and Proposition~\ref{prop:mse-bound} together give a procedure for choosing between a product ansatz and an entangling ansatz before any training.
\begin{enumerate}
\item \emph{Fourier transform the data.} From labelled samples of $f^*$, estimate the coefficients $c^*(\ell)$ on the encoding's frequency support $\Omega$, and group them into the tensors $C^{*(S)}$ with $|S| \geq 2$.
\item \emph{Measure the distance to rank one.} For each $C^{*(S)}$, compute $\rho_1(C^{*(S)})$, its squared distance to the nearest rank-one tensor. For $|S| = 2$, this is the sum of the squared singular values after the largest, Eq.~\eqref{eq:rho1-svd}. For $|S| \geq 3$, the same sum over a mode unfolding gives a lower bound.
\item \emph{Choose the architecture.} If $\rho_1(C^{*(S)}) = 0$ for every $S$, entangling gates are not necessary. Otherwise, the supports $S$ with $\rho_1(C^{*(S)}) > 0$ identify which qubits must be entangled, and $\sum_S \rho_1(C^{*(S)})$ is a lower bound on the MSE of leaving them unentangled.
\end{enumerate}
The diagnostic needs only a discrete Fourier transform of the dataset and singular value decompositions. Its verdict does not depend on which entangling gate is later chosen.

\section{\label{sec:rank}QFIM rank}
Two parameters can move the state along the same tangent direction, so the parameter count overstates the number of independent directions. Larocca et al.~\cite{larocca2023} relate the number of independent directions to the dynamical Lie algebra (DLA) of the ansatz. This section counts them with the QFIM rank. The count grows linearly with depth, at a rate $p_0$ set by the gate layout, until it reaches a ceiling $\Crank$ set by the DLA $\mathfrak g$ and the initial state $\ket{0}$.
The QFIM is the metric on the map from parameters to states. Its $(i,j)$ entry measures how much the output state changes when parameters $i$ and $j$ vary together.

\begin{definition}[QFIM]
For a state $\ket{\psi(\boldsymbol\theta)}$, the quantum Fisher information matrix $F$ has entries
\begin{equation}
F_{ij}(\boldsymbol\theta) := 4\,\mathrm{Re}\,\big\langle \partial_i\psi \,\big|\, (I - \ket{\psi}\bra\psi) \,\big|\, \partial_j\psi\big\rangle.
\end{equation}
\end{definition}

Its rank is the number of independent directions in which the parameters can move the state.

\begin{definition}[DLA]
\label{def:dla}
The dynamical Lie algebra $\mathfrak g$ of an ansatz is the real Lie algebra generated under nested commutators by the operators $iW^jH_kW^{-j}$, $j\in\Z$, where $H_k$ are the Pauli generators of the layer's rotations and $W$ is the layer's entangler, the product of its CX gates. We write $H\in\mathfrak g$ for $iH\in\mathfrak g$.
\end{definition}
Throughout this section, $U$ is a depth-$L$ circuit, $\ket\psi = U\ket0$ is its output state, and $F(L)$ is its QFIM. Its trainable rotations are $e^{-i\theta_\mu H_\mu/2}$, each generated by a Pauli string $H_\mu$, a tensor product of $I$, $X$, $Y$ and $Z$.
The QFIM rank is bounded above by $\Crank = \dim\mathfrak{g} - \dim\Stab_\mathfrak{g}(\ket{0})$. The stabiliser $\Stab_\mathfrak{g}(\ket{0}) = \{H \in \mathfrak{g} : H\ket{0} \propto \ket{0}\}$ is the subalgebra of generators that fix the initial state up to a phase. Sec.~\ref{sec:rank_orbit} derives this ceiling.
One layer does not in general reach $\Crank$. For a hardware-efficient circuit, Haug, Bharti, and Kim~\cite{haug2021} find numerically that the rank climbs linearly with depth until it saturates. We take the rate to be $p_0$ per layer, where $p_0$ is the number of independent rotation angles in one layer. Some generators of the first layer do not point in independent directions. We count them in a first-layer deficit $d_0 := p_0 - \max_{\boldsymbol\theta}\rank F(1)$, taken where one layer stays below the ceiling. We observe two sources of this degeneracy:
\begin{itemize}
    \item A generator $H_k$ may already fix the initial state up to a phase ($H_k\ket{0}\propto\ket{0}$). Its derivative is then parallel to $\ket{\psi}$ and the projector $(I-\ket{\psi}\bra\psi)$ removes it. The HVA chain shows this: its $N{-}1$ commuting $R_{ZZ}$ generators all fix $\ket{0}$, so at $N=2$ the lone $Z_0Z_1$ contributes $d_0 = 1$.
    \item A layer may have more on-site generators than the local tangent has room for. N-Local shows this: three rotations per qubit on a $2$-dimensional Bloch tangent give a deficit of one per qubit, so at $N=1$ the three rotations on $\ket{0}$ contribute $d_0 = 1$.
\end{itemize}
The rank law is therefore
\begin{equation}
\rank F(L) \leq \min\big(p_0\,L - d_0,\,\Crank\big).
\label{eq:rank_main}
\end{equation}
Table~\ref{tab:results} lists the ingredients of the law for $9$ families of common ans\"atze.

\subsection{\label{sec:rank_orbit}Orbit-stabiliser ceiling}

Every parameter derivative of the output state $\ket{\psi}$ lies in $\mathfrak g\ket{\psi}$, whose dimension is the same at every point of the orbit. The QFIM rank is therefore bounded by the dimension of the orbit of $\ket{0}$ at every depth. By the orbit-stabiliser theorem, the orbit of the state under $e^{\mathfrak g}$ has dimension $\dim\mathfrak g$ minus the dimension of the subalgebra that fixes it~\cite{albertini2003}, which gives
\begin{equation}
\Crank = \dim(\mathfrak g) - \dim\big(\Stab_{\mathfrak g}(\ket{0})\big).
\label{eq:rank_ceiling}
\end{equation}
As an example, take the hardware-efficient ansatz (HEA) on $N=3$ qubits with a CX chain. Single-qubit rotations span $\mathfrak{su}(2)$ on each site and the CX chain couples all qubits, so the DLA is the full $\mathfrak{su}(8)$ with $\dim\mathfrak g = 4^3 - 1 = 63$. The stabiliser of $\ket{000}$ inside $\mathfrak{su}(8)$ has dimension $63 - 2(2^3 - 1) = 49$. Its complement is the $14$-dimensional tangent to the space of pure states, $\mathbb{CP}^7$, so $\Crank = 14$. Each layer adds $p_0 = 2N = 6$ parameters, and there is no first-layer deficit ($d_0 = 0$). The rank grows as $6L$ until the ceiling, reached at the rank-saturation depth $\Lrank = \lceil 14/6\rceil = 3$. At $N=3$, layers beyond $L=3$ add parameters but no new directions.
When the qubits split into disconnected CX components, the state factorises, and the ceiling adds across components. A connected component of $q_b$ qubits with $\mathfrak{su}(2)$-spanning gates contributes $2(2^{q_b}-1)$, the real dimension of its projective pure-state manifold.
Haug, Bharti, and Kim~\cite{haug2021} bound the rank by $2^{N+1}-2$ for any ansatz, and Larocca et al.~\cite{larocca2023} by the orbit dimension, which is at most $\dim\mathfrak g$. Eq.~\eqref{eq:rank_ceiling} gives this dimension in closed form. For the HVA chain at $N=4$, where $\mathfrak g\cong\mathfrak{so}(8)$~\cite{wiersema2024}, the three bounds are $30$, $28$ and $19$, and Eq.~\eqref{eq:rank_main} gives the measured rank $4$ at $L=1$.

\subsection{\label{sec:rank_growth}Rank growth rate}
We now show why the rank grows by $p_0$ per layer: each new angle adds a direction that the earlier layers did not reach. The derivative of the output state with respect to $\theta_\mu$ is
\begin{equation}
\partial_\mu\ket\psi = -\tfrac{i}{2}\,U\hat H_\mu\ket0,
\label{eq:pullback}
\end{equation}
where $\hat H_\mu$ is $H_\mu$ conjugated by the gates before it~\cite{larocca2023}. Since $U$ is unitary, Eq.~\eqref{eq:pullback} shows that the QFIM is the Gram matrix of the vectors $(I-\ket0\bra0)\hat H_\mu\ket0$, and its rank is the dimension of their real span $V_L$. The bound in Eq.~\eqref{eq:rank_main} holds for every $\boldsymbol\theta$, and the rank is maximal for almost every $\boldsymbol\theta$, so a single parameter point that reaches the bound fixes the generic rank (Appendix~\ref{app:bounds}, Propositions~\ref{prop:upper} and~\ref{prop:generic}).

The count is easiest to see at $\theta_\mu = \pi/2$. Conjugating a Pauli string $P$ by a rotation $R = e^{-i\theta G/2}$ gives
\begin{equation}
R^\dagger P R =
\begin{cases}
P, & [P,G] = 0,\\
P\cos\theta + iGP\sin\theta, & \{P,G\} = 0.
\end{cases}
\label{eq:conjugation}
\end{equation}

At $\theta = \pi/2$ the second line of Eq.~\eqref{eq:conjugation} is the single string $iGP$. A CX also maps one string to one string, so every $\hat H_\mu$ is a single Pauli string. On $\ket{0}$, a Pauli string flips the qubits where it has $X$ or $Y$ and leaves the others alone, so each $\hat H_\mu\ket{0}$ is one basis state $\ket{x}$, up to a factor $\pm 1$ or $\pm i$. Counting directions then becomes counting bit strings: the rank at this point is the number of distinct $\ket{x}$ and $i\ket{x}$ with $x \neq 0$.

This picture shows why each layer adds new directions. A CX copies an $X$ on its control onto its target, so a generator conjugated by more layers becomes a longer string and flips more qubits. In EfficientSU2 at $N = 3$, the first layer flips one qubit at a time, giving states such as $\ket{100}$. The $R_y$ on $q_0$ in the second layer, conjugated by the first layer, becomes $Z_0 Y_1 Y_2$ and flips two qubits, giving $\ket{011}$, a state no first-layer angle reaches.

A special point such as $\pi/2$ can only lose directions, so this count is a lower bound on the generic rank. The exact ranks of Table~\ref{tab:results} are computed at random angles instead, $\theta_\mu = 4\arctan t_\mu$ with $t_\mu$ drawn uniformly from $\Z\cap(1,10^3]$ (seed $42$), so that every sine and cosine is rational and the rank is exact. The $\pi/2$ counting here is illustrative.

\begin{figure}[t]
\centering
\includegraphics[width=\columnwidth]{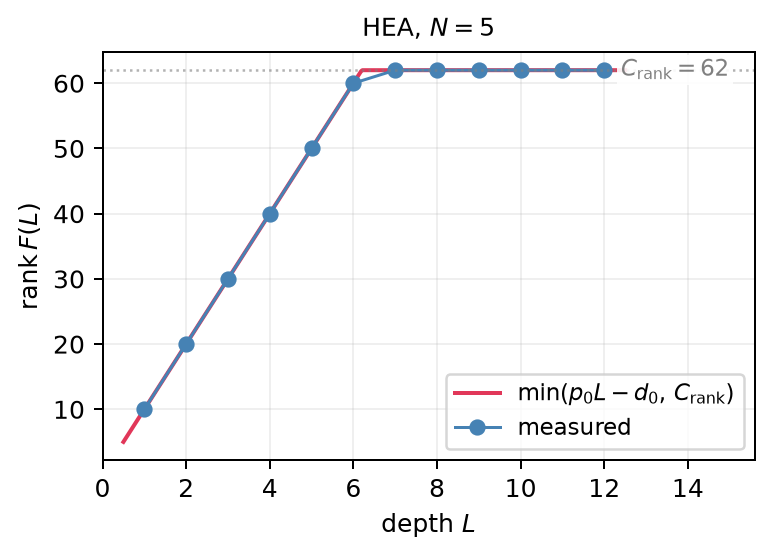}
\caption{The rank grows at slope $p_0$ and turns a sharp corner at the orbit-stabiliser ceiling. Markers: measured QFIM rank of HEA at $N=5$. Red line: the predictor $\min(p_0 L - d_0,\,\Crank)$.}
\label{fig:rank_rate_perfect}
\end{figure}

At every $(N,L)$ we have measured, the rate holds for generic $\boldsymbol\theta$: HEA, EfficientSU2 and N-Local at $N\leq12$ and every depth up to the ceiling, including the $s=3$ block of Sec.~\ref{sec:design}. An example for HEA is shown in Fig.~\ref{fig:rank_rate_perfect} at $N=5$: the measured rank follows a straight line of slope $p_0$ and turns a sharp corner at $\Crank$. We treat the rate as we would a full-rank Jacobian in a neural network. When it fails, it fails for every $\boldsymbol\theta$, so the cause lies in the circuit, and Sec.~\ref{sec:rank_frozen} shows how to find it from the gate list.

\subsection{\label{sec:rank_frozen}Frozen expectations}
The rate can fail because some Pauli expectation $\langle P\rangle$ stays at zero for every $\boldsymbol\theta$, which confines the reachable states to a smaller set. We find such $P$ by following the initial state forward through the gates. The initial state is the average of all $2^N$ Pauli-$Z$ products of Eq.~\eqref{eq:observable}, $\ket0\bra0 = 2^{-N}\sum_\alpha Z^\alpha$, so the expectation of any Pauli string $P$ in the output state is
\begin{equation}
\langle P\rangle = 2^{-N}\sum_\alpha \tr\big(P\,UZ^\alpha U^\dagger\big).
\label{eq:expectation}
\end{equation}
By Eq.~\eqref{eq:conjugation}, a rotation leaves a Pauli string unchanged or mixes it with one partner, while a CX maps one string to one string. We call the Pauli strings that $UZ^\alpha U^\dagger$ can contain the strings reachable from $Z^\alpha$. If no Pauli-$Z$ product reaches $P$, every term in Eq.~\eqref{eq:expectation} vanishes and $\langle P\rangle$ is frozen at zero for every $\boldsymbol\theta$. Each frozen expectation whose differential is independent on the tangent space of the orbit of $\ket0$ removes one direction from the ceiling, so
\begin{equation}
\rank F(L) \leq \Crank - \kappa(L),
\label{eq:frozen}
\end{equation}
where the codimension $\kappa(L)$ counts the independent frozen expectations at depth $L$. For the $\mathfrak{su}(2^N)$ families in Table~\ref{tab:results}, the first $R_yR_z$ already turns each $Z_q$ into $X_q$, $Y_q$ and $Z_q$, so every Pauli string is reachable after one layer, nothing is frozen, and Eq.~\eqref{eq:frozen} changes nothing.

In the HVA chain each layer extends the strings reachable from the end operators $Z_0$ and $Z_{N-1}$, which fix $\ket0$, by one site, so strings more than $L$ sites from the ends stay frozen. Appendix~\ref{app:hva} follows them and finds $4(N-L)$ frozen expectations, which lower the ceiling to $\Crank-4(N-L)$ once the two reachable sets overlap at $2L>N$. With Eq.~\eqref{eq:rank_main} this gives
\begin{equation}
\rank F(L) = \begin{cases}
p_0L{-}d_0, & 2L\leq N,\\
\Crank{-}4(N{-}L), & N<2L\leq2N,\\
\Crank, & L>N.
\end{cases}
\label{eq:hva_exact}
\end{equation}
The same steps apply to any ansatz: list the strings reachable from the Pauli-$Z$ products, count the strings that none of them reach, and check which of their differentials are independent.

\subsection{\label{sec:rank_validation}Validation}
We test the rank law against exact ranks of nine common ansatz families (Table~\ref{tab:results}). The HVA chain has $R_x$ on every site, so after exchanging $X$ and $Z$ its generators are those of the $\mathfrak{so}(2N)$ entry of Wiersema et al.~\cite{wiersema2024}. Their transverse-field Ising entry has no field on the first site and DLA $\mathfrak{so}(2N-1)$. The XY chain has DLA $\mathfrak{so}(2N)$ with or without the end fields.

Brickwork couples even pairs in one sublayer and odd pairs in the next, so we take one layer to be an even sublayer followed by an odd one, which connects all qubits. Each of its two end qubits then adds only two directions per layer, so $p_0 = 4N-4$ (worked through in Appendix~\ref{app:brick}). Two small-$N$ points miss the law, Brickwork at $N=4$, $L=2$ and the XY ring at $N=4$, $L=1$, and Appendix~\ref{app:smalln} explains both.

\begin{table*}[t]
\centering\small
\renewcommand{\arraystretch}{1.25}
\setlength{\tabcolsep}{2.6pt}
\caption{The rank law taken with equality, $\rank F(L) = \min(p_0 L - d_0,\,\Crank-\kappa(L))$, with $\Crank = \dim\mathfrak g - \dim\Stab_{\mathfrak g}(\ket{0})$ and the codimension $\kappa(L)$ of Sec.~\ref{sec:rank_frozen} ($0$ for $L>N$), predicts the measured rank exactly for every family at $N\in\{3,\ldots,6\}$ up to $L=12$, except at the two small-$N$ exceptions explained in Appendix~\ref{app:smalln}. Appendix~\ref{app:brick} works through the Brickwork row step by step.}
\label{tab:results}
\begin{tabular*}{\textwidth}{@{\extracolsep{\fill}}l l l l l r r l l@{}}
\toprule
Ansatz & Gate set & Coupling & DLA $\mathfrak g$ & $\dim\mathfrak g$ & $p_0$ & $d_0$ & $\Crank$ & $\kappa$ \\
\midrule
HEA~\cite{kandala2017}         & $R_yR_z$               & chain       & $\mathfrak{su}(2^N)$           & $4^N{-}1$       & $2N$       & $0$     & $2(2^N{-}1)$ & $0$ \\
EfficientSU2~\cite{qiskitlib}  & $R_yR_z$               & complete    & $\mathfrak{su}(2^N)$           & $4^N{-}1$       & $2N$       & $0$     & $2(2^N{-}1)$ & $0$ \\
Product                        & $R_yR_z$               & none        & $\mathfrak{su}(2)^N$           & $3N$            & $2N$       & $0$     & $2N$ & $0$ \\
N-Local~\cite{qiskitlib}       & $R_yR_zR_x$            & ring        & $\mathfrak{su}(2^N)$           & $4^N{-}1$       & $3N$       & $N$     & $2(2^N{-}1)$ & $0$ \\
HVA chain~\cite{wiersema2024}  & $R_{ZZ},R_x$           & chain       & $\mathfrak{so}(2N)$            & $N(2N{-}1)$     & $2N{-}1$   & $N{-}1$ & $N^2{+}N{-}1$ & $4(N{-}L)$ \\
HVA ring~\cite{wiersema2024}   & $R_{ZZ},R_x$           & ring        & $\mathfrak{so}(2N)^{\oplus 2}$ & $2N(2N{-}1)$    & $2N$       & $N$     & $2N^2{-}2N{+}1$ & $0$ \\
XY chain~\cite{wiersema2024}   & $R_{XX},R_{YY},R_z$    & chain       & $\mathfrak{so}(2N)$            & $N(2N{-}1)$     & $3N{-}2$   & $0$     & $N(N{-}1)$ & $0$ \\
XY ring~\cite{wiersema2024}    & $R_{XX},R_{YY},R_z$    & ring        & $\mathfrak{so}(2N)^{\oplus 2}$ & $2N(2N{-}1)$    & $3N$       & $0$     & $N(N{-}1)$ & $0$ \\
Brickwork~\cite{sim2019}       & $R_yR_z$               & even/odd    & $\mathfrak{su}(2^N)$           & $4^N{-}1$       & $4N{-}4$   & $N{-}\{3,4\}$ & $2(2^N{-}1)$ & $0$ \\
\bottomrule
\end{tabular*}
\end{table*}

\section{Extraction}
\label{sec:extraction}
The usual single-qubit Pauli-$Z$ readout resolves an exponentially shrinking fraction of the state. Reading out all Pauli-$Z$ products resolves half of it at every $N$, the most any single-basis readout can. The extraction ratio $\rho$ is this fraction: the share of an ansatz's reachable state directions that a readout resolves. Throughout, we restrict to pure states. Noisy circuits are outside the scope of this work.

An $N$-qubit pure state $\ket{\psi} \in \C^{2^N}$ has $2 \cdot 2^N$ real parameters. Normalisation $\|\psi\| = 1$ and the global phase symmetry $\ket{\psi} \sim e^{i\varphi}\ket{\psi}$ each remove one, leaving
\begin{equation}
\dim \manifold = 2(2^N - 1),
\label{eq:manifold_dim}
\end{equation}
the dimension of the projective pure-state manifold $\manifold$. Eq.~\eqref{eq:manifold_dim} is the largest number of independent real parameters any readout could resolve from a pure $N$-qubit state. The orbit-stabiliser ceiling $\Crank$ of Sec.~\ref{sec:rank} caps the dimension a given ansatz actually reaches, with $\Crank \leq \dim \manifold$.
A readout is a collection of Hermitian observables $\{O_1, \ldots, O_{\nobs}\}$ whose expectation values form the output of the VQC. We take $\nobs$ to be the number of linearly independent operators in the set, so that $\mathrm{span}_\R\{O_1, \ldots, O_{\nobs}\}$ has dimension $\nobs$ in the space of Hermitian matrices.
\begin{definition}[Extraction ratio]
\label{def:extraction_ratio}
For a readout $\{O_1, \ldots, O_{\nobs}\}$ acting on an ansatz with orbit-stabiliser ceiling $\Crank$,
\begin{equation}
\rho \;=\; \frac{\nobs}{\min(\Crank,\, \dim \manifold)}.
\label{eq:rho_def}
\end{equation}
\end{definition}
We call a readout \emph{single-basis} when a fixed orthonormal basis of $\C^{2^N}$ diagonalises all of its observables simultaneously. The expectation values of such a readout follow from the outcome statistics of one measurement basis, with no further post-rotations. A readout that additionally measures $X_q$ on some qubit is not single-basis, since $X_q$ and $Z_q$ are not simultaneously diagonalisable. Both readouts below are single-basis.
%
\subsection{Single-qubit readout}
\label{subsec:extr_single_qubit}
A common readout in variational quantum learning measures one observable per qubit, $\{Z_0, Z_1, \ldots, Z_{N-1}\}$, giving $\nobs = N$. On a fully expressive ansatz ($\Crank = \dim\manifold$), the extraction ratio is then
\begin{equation}
\rho \;=\; \frac{N}{2(2^N - 1)}.
\label{eq:rho_single}
\end{equation}
Eq.~\eqref{eq:rho_single} decays exponentially in $N$. At $N=3$ it is $3/14 \approx 21\%$, at $N=5$ it is $5/62 \approx 8\%$, and at $N=10$ it is already below $0.5\%$. The manifold grows exponentially with $N$, while $\nobs$ grows only linearly.

\subsection{Pauli-$Z$ products}
\label{subsec:extr_zprods}
Consider instead the readout of all non-trivial Pauli-$Z$ products $Z^\alpha$ of Eq.~\eqref{eq:observable}, one for every non-zero $\alpha \in \{0,1\}^N$. There are $2^N - 1$ such operators, and they are linearly independent as Hermitian matrices, so $\nobs = 2^N - 1$.
All $2^N - 1$ Pauli-$Z$ products follow from one batch of computational-basis shots. With $p_\mathbf{b} = |\braket{\mathbf{b}|\psi}|^2$ and $\alpha \cdot \mathbf{b} = \sum_q \alpha_q b_q$,
\begin{equation}
\braket{Z^\alpha} \;=\; \sum_{\mathbf{b} \in \{0,1\}^N} (-1)^{\alpha \cdot \mathbf{b}}\, p_\mathbf{b}.
\end{equation}
The matrix $H$ with entries $H_{\alpha,\mathbf{b}} = (-1)^{\alpha \cdot \mathbf{b}}$ is the Walsh-Hadamard matrix without its first row, so its rows are orthogonal and it has rank $2^N - 1$.
\begin{proposition}[Maximal single-basis extraction]
\label{prop:extraction_max}
On an ansatz saturating the pure-state manifold, the readout of all Pauli-$Z$ products achieves
\begin{equation}
\rho \;=\; \frac{2^N - 1}{2(2^N - 1)} \;=\; \frac{1}{2},
\label{eq:rho_zprods}
\end{equation}
independent of $N$. Moreover, no single-basis readout achieves $\rho > 1/2$.
\end{proposition}
The upper bound is a counting argument. A single orthonormal basis of $\C^{2^N}$ has $2^N$ outcomes and hence at most $2^N - 1$ independent probabilities, against the $2(2^N - 1)$ real parameters of a pure state~\cite{flammia2005}. Every single-basis expectation value is a linear functional of these probabilities, so $\nobs \leq 2^N - 1$. The factor of two between this count and $\dim \manifold = 2(2^N - 1)$ gives $\rho \leq 1/2$, and Eq.~\eqref{eq:rho_zprods} shows that all Pauli-$Z$ products attain it. Any $2^N - 1$ linearly independent diagonal observables attain it too, so the Pauli-$Z$ products are the natural choice, not the only one.
The missing half is the relative phases: one basis resolves the $2^N - 1$ independent moduli $|\braket{\mathbf{b}|\psi}|$ but not the $2^N - 1$ independent phases.

\section{\label{sec:sandwich}Depth and block size}
This section combines the QFIM rank law of Sec.~\ref{sec:rank} with a gradient-variance bound to fix the depth $L$ and the block size $s$. The rank ceiling fixes $L$ as a function of $s$. The gradient-variance bound (Sec.~\ref{sec:sandwich_grad}) gives an upper bound on $s$. At the parameter budget for the Madelon dataset, the two leave one admissible pair $(s, L)$.

\subsection{\label{sec:sandwich_phases}Depth}
Consider an HEA block of $s$ qubits: $\mathfrak{su}(2)$-spanning rotations on each qubit and a connected CX chain, so $p_0 = 2s$ and $d_0 = 0$. Eq.~\eqref{eq:rank_main} gives the depth at which the rank saturates,
\begin{equation}
\Lrank(s) = \left\lceil \tfrac{\Crank + d_0}{p_0} \right\rceil = \left\lceil \tfrac{2(2^s - 1)}{2s} \right\rceil = \left\lceil \tfrac{2^s - 1}{s} \right\rceil.
\label{eq:Lstar}
\end{equation}
We prescribe $L = \Lrank(s)$, the smallest depth at which the block has all $\Crank$ directions. At $s = 3$ this gives $L = 3$, and at $s = 5$ it gives $L = 7$.
\subsection{\label{sec:sandwich_grad}Block size}
Eq.~\eqref{eq:Lstar} fixes $L$ at a given block size $s$. The gradient magnitude bounds the block size itself from above. For a block of $s$ qubits whose layers form a 2-design, with input $\ket{0}$ and a Pauli-$Z$-product observable, the Lie-algebraic loss variance~\cite{larocca2025bp} and the gradient variance of McClean et al.~\cite{mcclean2018} (equivalently Fontana et al.~\cite{fontana2023}) evaluate to
\begin{equation}
\mathrm{Var}[C] = \frac{1}{2^s + 1}, \qquad \mathrm{Var}[\partial_\theta C] \approx 2^{-(s+1)},
\label{eq:block_var}
\end{equation}
both decaying as $2^{-s}$ because a Pauli-$Z$ product has $\mathrm{Tr}\,O^2 = 2^s$. At the depth $L = 3$ used here, the block is not yet a 2-design for $s \geq 4$, but the variances we measure over $1500$ random parameter draws are within a factor $1.7$ of Eq.~\eqref{eq:block_var} for $s = 2,\ldots,5$ and halve per added qubit. For a monolithic $N$-qubit circuit the same formulas give $2^{-N}$ decay, the barren-plateau regime~\cite{mcclean2018,larocca2025bp}.
Splitting the circuit into $K$ independent blocks of size $s$ removes the dependence on $N$. Each block's variance is fixed by $s$ and does not depend on $K$, while the total ceiling $\Crank = K\cdot 2(2^s-1)$ grows linearly in $K$.
Combining the prescribed depth with the gradient cap,
\begin{equation}
L = \Lrank(s), \qquad s\leq s_\mathrm{grad}(\epsilon_g),
\label{eq:sgrad_def}
\end{equation}
where $s_\mathrm{grad}(\epsilon_g)$ is the largest $s$ whose 2-design gradient variance $\mathrm{Var}[\partial_\theta C]$ of Eq.~\eqref{eq:block_var} is at least the trainability threshold $\epsilon_g$. For $\epsilon_g = 10^{-2}$, $s_\mathrm{grad} = 5$, and for $\epsilon_g = 10^{-3}$, $s_\mathrm{grad} = 8$. The cap binds at $s = 3$ only for $\epsilon_g\in(2.95\times10^{-2}, 5.64\times10^{-2}]$.
For the parameter budget on the Madelon dataset, we take $s = 3$ and $L = 3$. Then Eq.~\eqref{eq:block_var} gives $\mathrm{Var}[C] = 1/9$ and $\mathrm{Var}[\partial_\theta C] = 0.056$ (easily trainable).

\section{\label{sec:design}Block classifier}
The three considerations of Secs.~\ref{sec:function-class}, \ref{sec:rank} and \ref{sec:extraction} fix every hyperparameter of the block classifier.

\subsection{\label{sec:design_rules}Hyperparameters}
\textbf{Block size $s = 3$.} The gradient cap of Eq.~\eqref{eq:sgrad_def} binds at $s = 3$ for a trainability threshold $\epsilon_g = 5\times10^{-2}$, which we adopt. At the conventional barren-plateau threshold $\epsilon_g = 10^{-2}$ it allows $s\leq 5$, so $s = 3$ is then one of several admissible sizes. Sec.~\ref{sec:exp_ablation} compares $s = 2, 3, 4$ at the same parameter count.

\textbf{Depth $L = 3$.} Eq.~\eqref{eq:Lstar} gives $L = \lceil 7/3\rceil = 3$.

\textbf{Readout: Pauli-$Z$ products.} By Proposition~\ref{prop:extraction_max}, measuring all $2^s-1 = 7$ non-trivial Pauli-$Z$ products per block attains $\rho = 1/2$, the optimum for a single-basis readout.

\textbf{Number of blocks $K = 7$: set by the feature count.} The $K$ blocks must cover the $20$ selected features, so $K = \lceil 20/3\rceil = 7$. The one spare slot wraps around to the first feature. The parameter count is then an output: $3s$ parameters per layer, $L = 3$ layers and $2^s - 1 = 7$ readout weights per block, plus one bias, give $7\cdot(27 + 7) + 1 = 239$. The total ceiling is $K\cdot\Crank(s) = 98$ directions.

\textbf{Per-layer reuploading.} Each feature $x_q$ is encoded again as $R_y(w_{\ell q}\,x_q)$ at every layer $\ell$, with trainable $w_{\ell q}$. Reuploading widens the Fourier spectrum the circuit can represent~\cite{schuld2021}. The ablation in Sec.~\ref{sec:experiments} measures a $4.5$-point accuracy cost without it.

\subsection{\label{sec:design_circuit}The circuit}
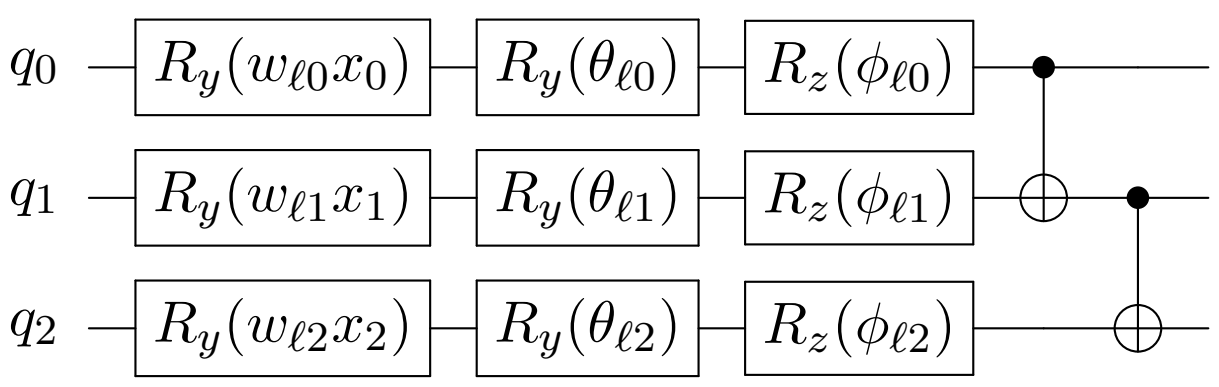
\begin{figure}[h]
\centering
$$
\Qcircuit @C=0.8em @R=0.6em {
\lstick{q_0} & \gate{R_y(w_{\ell 0} x_0)} & \gate{R_y(\theta_{\ell 0})} & \gate{R_z(\phi_{\ell 0})} & \ctrl{1} & \qw      & \qw \\
\lstick{q_1} & \gate{R_y(w_{\ell 1} x_1)} & \gate{R_y(\theta_{\ell 1})} & \gate{R_z(\phi_{\ell 1})} & \targ    & \ctrl{1} & \qw \\
\lstick{q_2} & \gate{R_y(w_{\ell 2} x_2)} & \gate{R_y(\theta_{\ell 2})} & \gate{R_z(\phi_{\ell 2})} & \qw      & \targ    & \qw
}
$$
\caption{One layer of a $3$-qubit block. Left: data encoding with trainable scale $w_{\ell q}$. Middle: variational $R_yR_z$. Right: CX chain. The layer repeats $L$ times with fresh encoding weights at each repetition (reuploading).}
\label{fig:block_layer}
\end{figure}

\begin{figure}[t]
\centering
\begin{adjustbox}{max width=\columnwidth}
\begin{tikzpicture}[
  font=\footnotesize,
  block/.style={draw, rounded corners=2pt, minimum width=1.9cm, minimum height=1.2cm, align=center, fill=blue!4},
  readout/.style={draw, rounded corners=2pt, minimum width=3.6cm, minimum height=0.8cm, align=center, fill=gray!8},
  arr/.style={-{Stealth[length=1.6mm]}, thick}
]
\node[font=\scriptsize] (i1) at (0,    1.5) {$x_0,\,x_1,\,x_2$};
\node[font=\scriptsize] (i2) at (2.6,  1.5) {$x_3,\,x_4,\,x_5$};
\node[font=\scriptsize] (iK) at (6.4,  1.5) {$\ldots,\,x_{3K-1}$};
\node[block] (b1) at (0,    0) {Block $1$ \\ HEA, depth $L$};
\node[block] (b2) at (2.6,  0) {Block $2$ \\ HEA, depth $L$};
\node                (cdots) at (4.5, 0) {$\cdots$};
\node[block] (bK) at (6.4,  0) {Block $K$ \\ HEA, depth $L$};
\draw[arr] (i1.south) to (b1.north);
\draw[arr] (i2.south) to (b2.north);
\draw[arr] (iK.south) to (bK.north);
\node[font=\scriptsize, align=center] (o1) at (0,   -1.5) {$2^s{-}1$ Pauli-$Z$\\products};
\node[font=\scriptsize, align=center] (o2) at (2.6, -1.5) {$2^s{-}1$ Pauli-$Z$\\products};
\node[font=\scriptsize, align=center] (oK) at (6.4, -1.5) {$2^s{-}1$ Pauli-$Z$\\products};
\draw[arr] (b1.south) to (o1.north);
\draw[arr] (b2.south) to (o2.north);
\draw[arr] (bK.south) to (oK.north);
\node[readout] (lr) at (3.2, -3.0) {linear readout $\mathbf w\cdot\mathbf f + b$};
\draw[arr] (o1.south) |- ([yshift=0pt]lr.west);
\draw[arr] (o2.south) -- ([xshift=-0.6cm]lr.north);
\draw[arr] (oK.south) |- ([yshift=0pt]lr.east);
\draw[arr] (lr.south) to ++(0,-0.5) node[below, font=\scriptsize] {logit};
\end{tikzpicture}
\end{adjustbox}
\vspace{-4pt}
\caption{The block classifier: $K$ blocks of $s$ qubits run in parallel on disjoint feature groups. The $2^s{-}1$ Pauli-$Z$ products of each block are concatenated into $\mathbf f$ and fed to a linear readout $\mathbf w\cdot\mathbf f + b$ that gives the logit.}
\label{fig:architecture}
\end{figure}
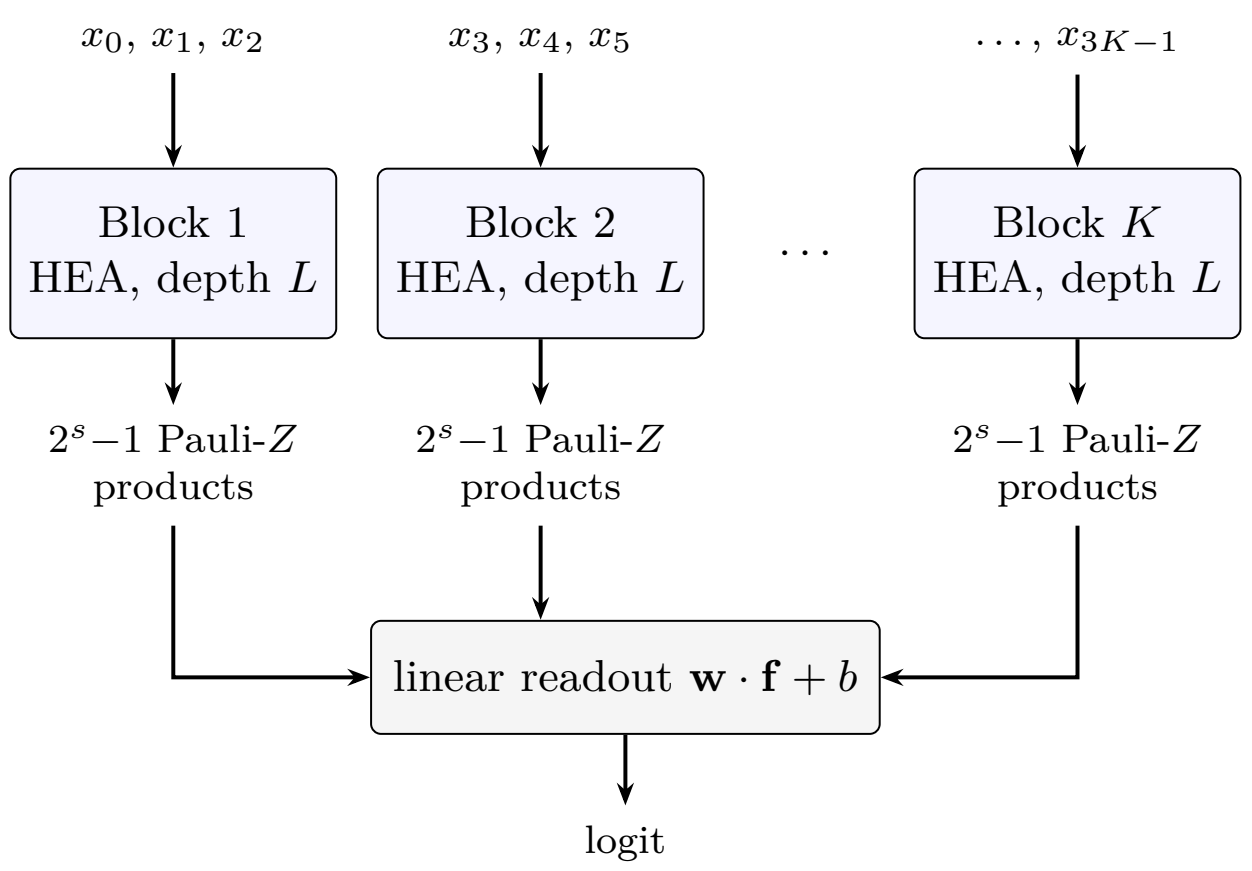

Each of the $K$ blocks repeats the layer of Fig.~\ref{fig:block_layer} $L$ times. One layer consists of $s$ data-encoding rotations with trainable scale, $s$ variational $R_y$, $s$ variational $R_z$, and a CX chain, for $3s$ parameters per layer. Only $2s$ of them count towards the QFIM rank. The encoding $R_y(w_{\ell q}x_q)$ and the variational $R_y(\theta_{\ell q})$ on the same qubit share an axis and merge into one rotation, so at a fixed input the tangent of $w_{\ell q}$ is $x_q$ times the tangent of $\theta_{\ell q}$. The rank law therefore uses $p_0 = 2s$, while the parameter count uses $3s$.
The full classifier runs $K$ such blocks in parallel on disjoint groups of features (Fig.~\ref{fig:architecture}). The $2^s-1$ Pauli-$Z$ products of each block are concatenated into a feature vector $\mathbf f$ of length $K(2^s-1)$, and a linear readout $\mathbf w\cdot\mathbf f + b$ gives the classification logit.

\section{\label{sec:experiments}Experiments}
\subsection{\label{sec:exp_data}Dataset and training}
We use the Madelon dataset~\cite{madelon2004} from the NIPS 2003 Feature Selection Challenge: $2600$ labelled samples and $500$ features, of which $5$ are informative, $15$ are redundant linear combinations of them, and $480$ are noise. Points are drawn from $32$ clusters at the vertices of a five-dimensional hypercube, each cluster with a random class label. Linear models fail: logistic regression reaches $61.5\%$ and a single-neuron MLP is random. We keep the $20$ features with the highest $F$-score against the label. The retained set contains all $5$ informative and all $15$ redundant features. We use an $80/20$ stratified train/test split.
We train all models with AdamW and cosine learning-rate decay (initial learning rate $0.005$), batch size $64$, for $200$ epochs. We run each configuration with five random seeds $(42,123,456,789,1337)$. Table~\ref{tab:ablation} reports the mean and standard deviation over seeds.

\subsection{\label{sec:exp_ablation}Ablations}
The baseline is the prescribed block classifier: $K = 7$ blocks of $s = 3$ qubits, depth $L = 3$, Pauli-$Z$-product readout, per-layer reuploading. It reaches a mean accuracy of $78.5\%$ with $239$ parameters. Each ablation changes a single design element. Three of the expected effects in Table~\ref{tab:ablation} are absent on the Madelon dataset, and the paragraphs below say which.
\begin{table*}[t]
\centering\small
\caption{Ablations, grouped by the consideration each one stresses. Three of the expected effects are absent. Accuracy is the mean and standard deviation over $5$ seeds. $\Delta$ is the difference of means from the baseline in percentage points.}
\label{tab:ablation}
\begin{tabular}{@{}p{5.5cm}rrrp{5.5cm}@{}}
\toprule
Ablation & Params & Acc (\%) & $\Delta$ & Expected from the consideration \\
\midrule
\raggedright \textbf{Baseline} ($K = 7$, $s = 3$, $L=3$, reuploading, Pauli-$Z$ products) & \textbf{239} & $\mathbf{78.5} \pm 1.8$ & n/a & \raggedright n/a \tabularnewline
\midrule
\multicolumn{5}{@{}l}{\textit{QFIM rank: depth sweep around $\Lrank(3) = 3$}} \\
\raggedright \quad $L = 1$ (rank at $43\%$ of ceiling)  & 113 & $74.6 \pm 0.9$ & $-3.9$ & \raggedright Lower: block rank $6$ vs.\ ceiling $14$ \tabularnewline
\raggedright \quad $L = 2$ (rank at $86\%$)              & 176 & $75.6 \pm 1.4$ & $-2.8$ & \raggedright Lower: block rank $12$ vs.\ $14$ \tabularnewline
\raggedright \quad $L = 5$ (rank saturated, $126$ extra parameters)  & 365 & $81.9 \pm 1.0$ & $+3.5$ & \raggedright No gain: no new QFIM directions past $\Lrank$ (not observed) \tabularnewline
\raggedright \quad $L = 8$ (rank saturated, $315$ extra parameters)  & 554 & $82.2 \pm 1.7$ & $+3.7$ & \raggedright Same (not observed) \tabularnewline
\midrule
\multicolumn{5}{@{}l}{\textit{QFIM rank: block size at a matched budget of $\sim 240$ parameters}} \\
\raggedright \quad $s = 2$, $K = 16$                    & 241 & $74.1 \pm 2.2$ & $-4.4$ & \raggedright Lower: $3$ features per block; $2.3\times$ more blocks needed \tabularnewline
\raggedright \quad $s = 4$, $K = 4$                      & 253 & $78.5 \pm 1.1$ & $0.0$ & \raggedright Lower: loss variance $1/17$ vs.\ $1/9$, Eq.~\eqref{eq:block_var} (not observed) \tabularnewline
\midrule
\multicolumn{5}{@{}l}{\textit{Extraction}} \\
\raggedright \quad Single-qubit $\langle Z_q\rangle$ only & 211 & $78.5 \pm 0.8$ & $0.0$ & \raggedright Lower: $\rho$ drops from $1/2$ to $3/14$, Eq.~\eqref{eq:rho_def} (not observed) \tabularnewline
\midrule
\multicolumn{5}{@{}l}{\textit{Reuploading}} \\
\raggedright \quad No reuploading (encode once)          & 176 & $74.0 \pm 1.3$ & $-4.5$ & \raggedright Lower: Fourier spectrum limited to $\{-1,0,1\}^s$ per feature \tabularnewline
\bottomrule
\end{tabular}
\end{table*}
\paragraph{Depth.} Below $\Lrank$ the accuracy rises with depth, as the rank law expects. Past $\Lrank$ it does not plateau: $L = 5$ gains $3.5$ points and $L = 8$ gains $3.7$, so almost all of the further gain lies between $L = 3$ and $L = 5$. The rank law predicts no new directions past $\Lrank$. Overparameterisation past rank saturation is known to ease optimisation~\cite{larocca2023}, which is one candidate explanation. We have not tested it.

\paragraph{Block size.} At a matched budget of $\sim 240$ parameters, $s = 3$ beats $s = 2$ by $4.4$ points, because an $s = 2$ block produces only $3$ features. The expected penalty at $s = 4$ (loss variance $1/17$ against $1/9$, Eq.~\eqref{eq:block_var}) does not appear, which agrees with the gradient cap at $\epsilon_g = 10^{-2}$ that allows $s\leq 5$ (Sec.~\ref{sec:sandwich_grad}). Halving the loss variance costs nothing measurable on this dataset.

\paragraph{Readout.} Dropping the Pauli-$Z$ products and keeping only the single-qubit $\langle Z_q\rangle$ leaves the mean accuracy unchanged, although $\rho$ falls from $1/2$ to $3/14$ (Eq.~\eqref{eq:rho_def}). The extraction ratio counts the state directions a readout can resolve, not the directions the labels depend on. On this target, the three single-qubit expectations per block hold as much usable information as the seven products.

\paragraph{Reuploading.} With the data encoded once instead of at every layer, accuracy drops by $4.5$ points, the largest effect in the table.

\section{\label{sec:conclusion}Conclusion}
We propose three rules to fix a VQC architecture before training. A product ansatz has a rank-one Fourier coefficient tensor on every subset of qubits (Theorem~\ref{thm:rank-one-constraint}). The singular values of a dataset's Fourier coefficients therefore decide whether entanglement is needed and bound the error of leaving it out (Proposition~\ref{prop:mse-bound}). The QFIM rank grows as $p_0 L - d_0$ until the ceiling $\Crank$ (Eq.~\eqref{eq:rank_main}), which fixes the depth. Reading out all Pauli-$Z$ products gives $\rho = 1/2$, the most any single-basis readout can (Proposition~\ref{prop:extraction_max}).

Ablations of the prescribed classifier confirm three predicted effects: shallower circuits, smaller blocks and dropping reuploading all lower accuracy. Deeper circuits still gain, because reuploading adds input frequencies that the rank law does not count.

We recognise a few open problems. The rules assume pure states, and noise changes the counting behind $\Crank$ and $\rho$. Mid-circuit measurements change both the state and the readout. A rank law that includes reuploading would also predict the gain past the saturation depth.

\section*{Acknowledgements}
We would like to thank the Mphasis F1 Foundation for their financial support.

\section*{Author contributions}
[Contributions of M.S., A.D., and A.P.] Large language models (Claude, Anthropic) and Grammarly were used for grammar checking and language editing, bibliographic search, checking code and calculations. The authors checked all results and are fully responsible for the work.

\bibliographystyle{quantum}
\bibliography{ref}

\begin{thebibliography}{10}

\bibitem{martyniuk2024qas}
Darya Martyniuk, Johannes Jung, and Adrian Paschke.
\newblock ``Quantum architecture search: a survey''.
\newblock In 2024 IEEE International Conference on Quantum Computing and Engineering (QCE).
\newblock \href{https://dx.doi.org/10.1109/QCE60285.2024.00198}{Pages 1695--1706}.
\newblock ~(2024).

\bibitem{sim2019}
Sukin Sim, Peter~D. Johnson, and Al{\'{a}}n Aspuru-Guzik.
\newblock ``Expressibility and entangling capability of parameterized quantum circuits for hybrid quantum-classical algorithms''.
\newblock \href{https://dx.doi.org/10.1002/qute.201900070}{Advanced Quantum Technologies {\bf 2}, 1900070}~(2019).

\bibitem{abbas2021}
Amira Abbas, David Sutter, Christa Zoufal, Aur{\'{e}}lien Lucchi, Alessio Figalli, and Stefan Woerner.
\newblock ``The power of quantum neural networks''.
\newblock \href{https://dx.doi.org/10.1038/s43588-021-00084-1}{Nature Computational Science {\bf 1}, 403--409}~(2021).

\bibitem{bowles2024}
Joseph Bowles, Shahnawaz Ahmed, and Maria Schuld.
\newblock ``Better than classical? {The} subtle art of benchmarking quantum machine learning models''~(2024).
\newblock  \href{http://arxiv.org/abs/2403.07059}{arXiv:2403.07059}.

\bibitem{schuld2021}
Maria Schuld, Ryan Sweke, and Johannes~Jakob Meyer.
\newblock ``Effect of data encoding on the expressive power of variational quantum machine learning models''.
\newblock \href{https://dx.doi.org/10.1103/PhysRevA.103.032430}{Physical Review A {\bf 103}, 032430}~(2021).

\bibitem{casas2023}
Berta Casas and Alba Cervera-Lierta.
\newblock ``Multidimensional {Fourier} series with quantum circuits''.
\newblock \href{https://dx.doi.org/10.1103/PhysRevA.107.062612}{Physical Review A {\bf 107}, 062612}~(2023).

\bibitem{wiedmann2025}
Marco Wiedmann, Maniraman Periyasamy, and Daniel~D. Scherer.
\newblock ``Fourier analysis of variational quantum circuits for supervised learning''.
\newblock In 2025 IEEE International Conference on Quantum Computing and Engineering (QCE).
\newblock \href{https://dx.doi.org/10.1109/QCE65121.2025.00196}{Pages 1785--1795}.
\newblock ~(2025).

\bibitem{kolda2009}
Tamara~G. Kolda and Brett~W. Bader.
\newblock ``Tensor decompositions and applications''.
\newblock \href{https://dx.doi.org/10.1137/07070111X}{SIAM Review {\bf 51}, 455--500}~(2009).

\bibitem{singh2008}
Ajit~P. Singh and Geoffrey~J. Gordon.
\newblock ``Relational learning via collective matrix factorization''.
\newblock In Proceedings of the 14th ACM SIGKDD International Conference on Knowledge Discovery and Data Mining.
\newblock \href{https://dx.doi.org/10.1145/1401890.1401969}{Pages 650--658}.
\newblock ~(2008).

\bibitem{larocca2023}
Mart{\'{i}}n Larocca, Nathan Ju, Diego Garc{\'{i}}a-Mart{\'{i}}n, Patrick~J. Coles, and Marco Cerezo.
\newblock ``Theory of overparametrization in quantum neural networks''.
\newblock \href{https://dx.doi.org/10.1038/s43588-023-00467-6}{Nature Computational Science {\bf 3}, 542--551}~(2023).

\bibitem{haug2021}
Tobias Haug, Kishor Bharti, and M.~S. Kim.
\newblock ``Capacity and quantum geometry of parametrized quantum circuits''.
\newblock \href{https://dx.doi.org/10.1103/PRXQuantum.2.040309}{PRX Quantum {\bf 2}, 040309}~(2021).

\bibitem{albertini2003}
Francesca Albertini and Domenico D'Alessandro.
\newblock ``Notions of controllability for bilinear multilevel quantum systems''.
\newblock \href{https://dx.doi.org/10.1109/TAC.2003.815027}{IEEE Transactions on Automatic Control {\bf 48}, 1399--1403}~(2003).

\bibitem{wiersema2024}
Roeland Wiersema, Efekan K{\"{o}}kc{\"{u}}, Alexander~F. Kemper, and Bojko~N. Bakalov.
\newblock ``Classification of dynamical {Lie} algebras of 2-local spin systems on linear, circular and fully connected topologies''.
\newblock \href{https://dx.doi.org/10.1038/s41534-024-00900-2}{npj Quantum Information {\bf 10}, 110}~(2024).

\bibitem{kandala2017}
Abhinav Kandala, Antonio Mezzacapo, Kristan Temme, Maika Takita, Markus Brink, Jerry~M. Chow, and Jay~M. Gambetta.
\newblock ``Hardware-efficient variational quantum eigensolver for small molecules and quantum magnets''.
\newblock \href{https://dx.doi.org/10.1038/nature23879}{Nature {\bf 549}, 242--246}~(2017).

\bibitem{qiskitlib}
{IBM Quantum}.
\newblock ``Circuit library: {N}-local circuits''~(2026).
\newblock Accessed 2 November 2026.

\bibitem{flammia2005}
Steven~T. Flammia, Andrew Silberfarb, and Carlton~M. Caves.
\newblock ``Minimal informationally complete measurements for pure states''.
\newblock \href{https://dx.doi.org/10.1007/s10701-005-8658-z}{Foundations of Physics {\bf 35}, 1985--2006}~(2005).

\bibitem{larocca2025bp}
Mart{\'{i}}n Larocca, Supanut Thanasilp, Samson Wang, Kunal Sharma, Jacob Biamonte, Patrick~J. Coles, Lukasz Cincio, Jarrod~R. McClean, Zo{\"{e}} Holmes, and M.~Cerezo.
\newblock ``Barren plateaus in variational quantum computing''.
\newblock \href{https://dx.doi.org/10.1038/s42254-025-00813-9}{Nature Reviews Physics {\bf 7}, 174--189}~(2025).

\bibitem{mcclean2018}
Jarrod~R. McClean, Sergio Boixo, Vadim~N. Smelyanskiy, Ryan Babbush, and Hartmut Neven.
\newblock ``Barren plateaus in quantum neural network training landscapes''.
\newblock \href{https://dx.doi.org/10.1038/s41467-018-07090-4}{Nature Communications {\bf 9}, 4812}~(2018).

\bibitem{fontana2023}
Enrico Fontana, Dylan Herman, Shouvanik Chakrabarti, Niraj Kumar, Romina Yalovetzky, Jamie Heredge, Shree~Hari Sureshbabu, and Marco Pistoia.
\newblock ``Characterizing barren plateaus in quantum ans{\"{a}}tze with the adjoint representation''.
\newblock \href{https://dx.doi.org/10.1038/s41467-024-49910-w}{Nature Communications {\bf 15}, 7171}~(2024).

\bibitem{madelon2004}
Isabelle Guyon.
\newblock ``Madelon''.
\newblock {UCI} Machine Learning Repository~(2004).

\bibitem{mityagin2020}
Boris~S. Mityagin.
\newblock ``The zero set of a real analytic function''.
\newblock \href{https://dx.doi.org/10.1134/S0001434620030189}{Mathematical Notes {\bf 107}, 529--530}~(2020).

\end{thebibliography}

\appendix

\section{\label{app:bounds}Rank bounds}
\begin{proposition}[upper bound]\label{prop:upper}
For every $\boldsymbol\theta$ and every $L\geq1$, $\dim V_L \leq \min(p_0 L - d_0,\,\Crank)$.
\end{proposition}
\begin{proof}
Every $\hat H_\mu$ lies in $\mathfrak g$, so $V_L$ lies in the tangent space of the orbit of $\ket0$, whose dimension is $\Crank$ by Eq.~\eqref{eq:rank_ceiling}. By the definition of $d_0$ the first layer gives at most $p_0 - d_0$ directions, and every further layer adds at most one direction per independent rotation, which is $p_0$. Together these give the bound.
\end{proof}
\begin{proposition}[one point suffices]\label{prop:generic}
The rank of $F(L)$ takes its maximum value at almost every $\boldsymbol\theta$. If one point reaches the bound of Proposition~\ref{prop:upper}, equality holds at almost every point.
\end{proposition}
\begin{proof}
Let $r$ be the maximum rank. An $r\times r$ minor that is non-zero at one point is real-analytic in $\boldsymbol\theta$ and not identically zero, so it vanishes only on a set of measure zero~\cite{mityagin2020}, and off that set the rank is $r$.
\end{proof}

\section{\label{app:hva}HVA chain}
\begin{figure}[t]
\centering
\begin{tikzpicture}[x=7.4mm,y=-6.2mm,font=\scriptsize,
  rot/.style={draw,fill=white,minimum width=5.6mm,minimum height=3.4mm,inner sep=1pt}]
\foreach \q in {0,...,3}{ \draw (0.3,\q) -- (6.7,\q); \node[left] at (0.3,\q) {$q_\q$}; }
\foreach \x/\a in {1/0,1/2,2/1,4/0,4/2,5/1}{
  \draw[fill=white] (\x-0.38,\a-0.28) rectangle (\x+0.38,\a+1.28);
  \node at (\x,\a+0.5) {$R_{ZZ}$};
}
\foreach \x in {3,6} \foreach \q in {0,...,3} \node[rot] at (\x,\q) {$R_x$};
\draw[decorate,decoration={brace,mirror,amplitude=3pt}] (0.6,3.45) -- (3.4,3.45) node[midway,below=3pt] {layer $1$};
\draw[decorate,decoration={brace,mirror,amplitude=3pt}] (3.6,3.45) -- (6.4,3.45) node[midway,below=3pt] {layer $2$};
\end{tikzpicture}
\caption{Two layers of the HVA chain at $N=4$. Each layer applies $R_{ZZ}$ on every bond and then $R_x$ on every qubit, so conjugation moves $Z_0$ by one qubit per layer.}
\label{fig:hva}
\end{figure}
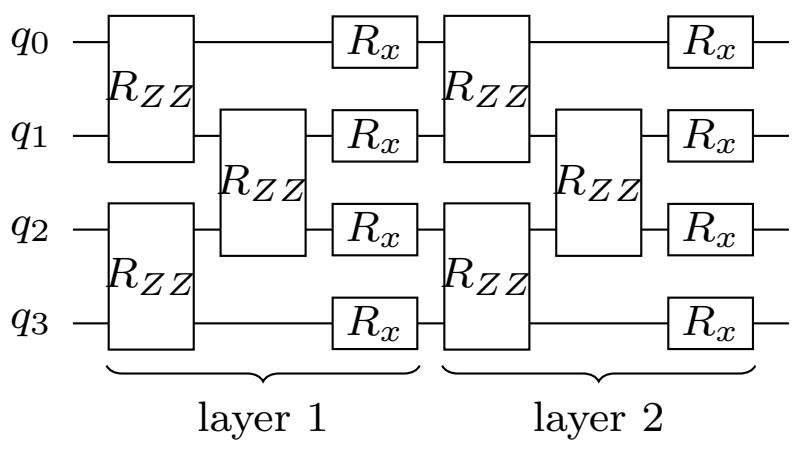

This appendix derives the middle case of Eq.~\eqref{eq:hva_exact}, where the rank grows by only $4$ per layer. A layer applies $R_{ZZ}$ to every bond $(j,j+1)$ and then $R_x$ to every qubit (Fig.~\ref{fig:hva}). This slowdown comes from the end operators $Z_0$ and $Z_{N-1}$, which fix $\ket{0}$. After $L$ layers, $Z_0$ conjugated by the circuit acts only on the first $L$ qubits. Similarly for $Z_{N-1}$, which acts the same way on the last $L$ qubits.

\textit{Reachable strings}: Conjugating $Z_0$ gate by gate with Eq.~\eqref{eq:conjugation} adds one string at each step:
\begin{itemize}
    \item $R_x$ on $q_0$ mixes $Z_0$ with $S_1=Y_0$.
    \item $R_{ZZ}$ on $(0,1)$ mixes $Y_0$ with $S_2=X_0Z_1$.
    \item $R_x$ on $q_1$ mixes $X_0Z_1$ with $S_3=X_0Y_1$.
\end{itemize}
Continuing along the chain gives the $2N$ strings
\begin{equation}
S_{2j} = X_0\cdots X_{j-1}Z_j,\ S_{2j+1} = X_0\cdots X_{j-1}Y_j,
\label{eq:hva_strings}
\end{equation}
with $j=0,\ldots,N-1$. Each gate mixes only neighbours in this list. One layer therefore moves $Z_0$ two places down the list, and after $L$ layers it reaches only $S_0,\ldots,S_{2L-1}$.

\textit{Frozen values}: Each gate maps the strings of Eq.~\eqref{eq:hva_strings} among themselves, so no string outside the list ever enters it. $Z_0$ is the only Pauli-$Z$ product in the list, so by Eq.~\eqref{eq:expectation}, at every $\boldsymbol\theta$,
\begin{equation}
\langle S_a\rangle = 0, \qquad a\geq 2L.
\label{eq:hva_frozen}
\end{equation}
The same holds from $Z_{N-1}$, so Eq.~\eqref{eq:hva_frozen} and its mirror freeze $4(N-L)$ values.

\section{\label{app:brick}Worked example}
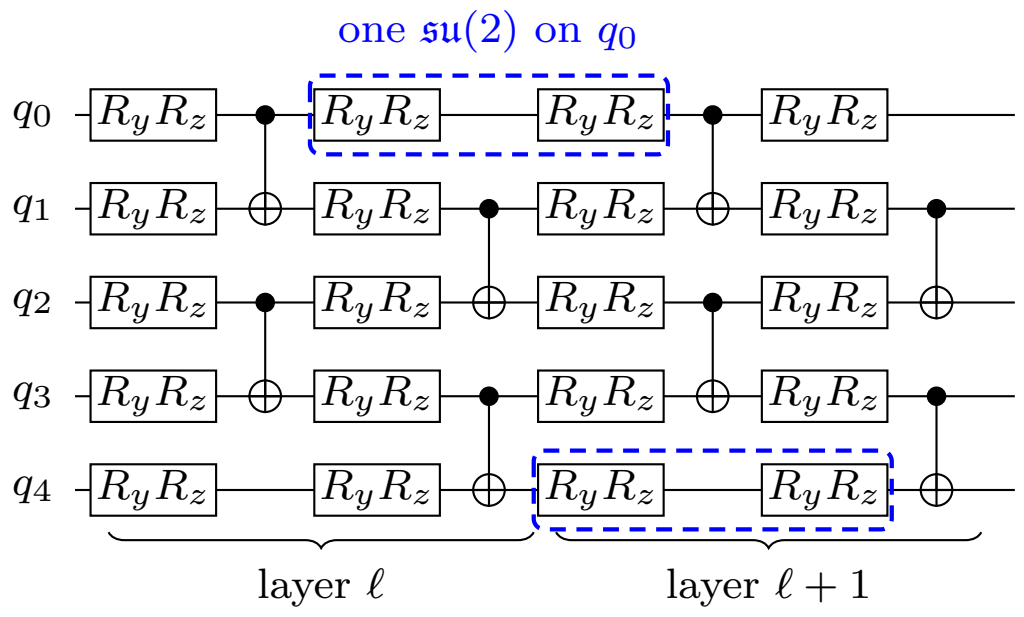
\begin{figure}[t]
\centering
\begin{tikzpicture}[x=7.4mm,y=-6.2mm,font=\scriptsize,
  rot/.style={draw,fill=white,minimum width=5.6mm,minimum height=3.4mm,inner sep=1pt},
  ctrl/.style={circle,fill,inner sep=1.3pt},
  targ/.style={circle,draw,inner sep=2.1pt}]
\foreach \q in {0,...,4}{ \draw (0.3,\q) -- (8.7,\q); \node[left] at (0.3,\q) {$q_\q$}; }
\foreach \x in {1,3,5,7} \foreach \q in {0,...,4} \node[rot] at (\x,\q) {$R_yR_z$};
\foreach \x in {2,6}{
  \draw (\x,0) -- (\x,1); \node[ctrl] at (\x,0) {}; \node[targ] (t) at (\x,1) {}; \draw (t.north) -- (t.south) (t.west) -- (t.east);
  \draw (\x,2) -- (\x,3); \node[ctrl] at (\x,2) {}; \node[targ] (t) at (\x,3) {}; \draw (t.north) -- (t.south) (t.west) -- (t.east);
}
\foreach \x in {4,8}{
  \draw (\x,1) -- (\x,2); \node[ctrl] at (\x,1) {}; \node[targ] (t) at (\x,2) {}; \draw (t.north) -- (t.south) (t.west) -- (t.east);
  \draw (\x,3) -- (\x,4); \node[ctrl] at (\x,3) {}; \node[targ] (t) at (\x,4) {}; \draw (t.north) -- (t.south) (t.west) -- (t.east);
}
\draw[blue,thick,densely dashed,rounded corners=2pt] (2.4,-0.42) rectangle (5.6,0.42);
\draw[blue,thick,densely dashed,rounded corners=2pt] (4.4,3.58) rectangle (7.6,4.42);
\node[blue,above] at (4,-0.45) {one $\mathfrak{su}(2)$ on $q_0$};
\draw[decorate,decoration={brace,mirror,amplitude=3pt}] (0.6,4.45) -- (4.4,4.45) node[midway,below=3pt] {layer $\ell$};
\draw[decorate,decoration={brace,mirror,amplitude=3pt}] (4.6,4.45) -- (8.4,4.45) node[midway,below=3pt] {layer $\ell+1$};
\end{tikzpicture}
\caption{Two Brickwork layers at $N=5$, each an even sublayer followed by an odd one. The end qubits $q_0$ and $q_4$ meet one CX per layer, so their four rotations between two CXs (dashed) form one $\mathfrak{su}(2)$ and add two directions per layer, not four.}
\label{fig:brick}
\end{figure}

This appendix applies the rank law step by step to Brickwork at $N=5$, as an illustrative example. Fig.~\ref{fig:brick} shows two of its layers. Each layer applies $R_yR_z$ to every qubit, then CX on the pairs $(0,1)$ and $(2,3)$, then $R_yR_z$ again, then CX on $(1,2)$ and $(3,4)$, which makes $20$ angles per layer.

\textit{Rate $p_0$}: A layer has $20$ angles but adds only $16$ directions. Two rules decide which rotation gates add a direction:
\begin{enumerate}[label=\textbf{Rule \arabic*.},leftmargin=*]
    \item Rotation gates on a qubit between two CXs add at most three directions.
    \item $R_z$ on a CX control and $R_x$ on a CX target commute with the CX, so they add no new direction.
\end{enumerate}
Per layer, this gives:
\begin{itemize}
    \item $4$ directions for each middle qubit $q_1$, $q_2$ and $q_3$. A CX acts on each of them in both sublayers, so between consecutive CXs it has two effective rotation gates, due to Rule 1.
    \item $2$ directions for the end qubit $q_0$. Only the CX on $(0,1)$ acts on it, so between consecutive CXs it has four rotation gates. By Rule 1 they combine into one $R_zR_yR_z$, and by Rule 2 the last $R_z$ adds no new direction.
    \item Similarly, $2$ directions for $q_4$, with $R_x$ in place of $R_z$.
\end{itemize}
One layer therefore adds $p_0 = 4+4+4+2+2 = 16$ directions.

\textit{Deficit $d_0$}: In the first layer the CX gates on $(1,2)$ and $(3,4)$ come last. As fixed gates they do not change the rank, and before them the qubits form three separate groups: $q_0q_1$, $q_2q_3$ and $q_4$ on its own. A pair of qubits has at most $2(2^2-1)=6$ directions and a single qubit $2$, so the first layer reaches at most $6+6+2=14$ directions, and the exact rank is $14$. This gives $d_0 = 2$. For any $N$ the first layer splits the same way, into pairs plus one lone qubit when $N$ is odd, which gives
\begin{equation}
d_0 =
\begin{cases}
N-3, & N \text{ odd},\\
N-4, & N \text{ even}.
\end{cases}
\label{eq:brick_d0}
\end{equation}
Eq.~\eqref{eq:brick_d0} is the $d_0$ entry of Table~\ref{tab:results}.

The ceiling follows from the DLA. The rotations span $\mathfrak{su}(2)$ on every qubit and the CX gates connect all five, so $\mathfrak g = \mathfrak{su}(32)$, of dimension $4^5-1 = 1023$. A generator fixes $\ket{0}$ when its first column is zero below the top entry, which takes $2(2^5-1) = 62$ real conditions, so $\Crank = 62$. Nothing is frozen either, since the first $R_yR_z$ already turns each $Z_q$ into $X_q$, $Y_q$ and $Z_q$.

Putting the three numbers into Eq.~\eqref{eq:rank_main} gives
\begin{equation}
\rank F(L) = \min(16L-2,\,62),
\label{eq:brick_example}
\end{equation}
so the rank should grow as $14$, $30$, $46$ and reach the ceiling at $\Lrank = 4$. The exact ranks are $14$, $30$, $46$, $62$, $62$ and $62$ for $L=1,\ldots,6$, as Eq.~\eqref{eq:brick_example} predicts.

\section{\label{app:smalln}Small-$N$ exceptions}
Two measured points miss the law of Table~\ref{tab:results}. In both, part of the circuit reaches fewer states than its angles count.

\textbf{XY ring, $N=4$, $L=1$.} Its $R_{XX}$ and $R_{YY}$ gates have $8$ angles but reach only $6$ directions, because $6$ independent expectations stay frozen at zero (Sec.~\ref{sec:rank_frozen}). The four $R_z$ gates add at most four more, so the rank is $10$, not $12$. At $N=5$ and $6$ every angle of these gates adds a direction, and the law holds.

\textbf{Brickwork, $N=4$, $L=2$.} Two layers have six CX gates. The last one acts after every rotation gate, so it does not change the rank. Of the other five, only one acts across the cut between $\{q_0,q_1\}$ and $\{q_2,q_3\}$, and one CX at most doubles the Schmidt rank across a cut. The state therefore has Schmidt rank at most $2$ across the cut. A $4\times4$ matrix of rank $2$ has $2(4+4-2)=12$ complex parameters, and removing norm and phase leaves $22$ real ones, so the rank is $22$, not $24$. At $L=3$ a second CX acts across the cut, and the rank reaches $\Crank=30$.

\end{document}